\documentclass[showpacs,onecolumn,floats,superscriptaddress]{revtex4}
\usepackage{amssymb}

\usepackage{graphicx}

\begin{document}

\title{Odd triplet superconductivity in superconductor-ferromagnet structure
with a narrow domain wall. }
\author{A.F. Volkov$^{1,2}$, K. B. Efetov$^{3}$.}

\address{Theoretische Physik III,\\
Ruhr-Universit\"{a}t Bochum, D-44780 Bochum, Germany\\
$^{(2)}$Institute for Radioengineering and Electronics of Russian Academy of\\
Sciences,11-7 Mokhovaya str., Moscow 125009, Russia\\
$^{(3)}$L.~D.~Landau Institute for Theoretical Physics RAS, 119334
Moscow, Russia}

\begin{abstract}
We study proximity effect in superconductor-ferromagnet (SF) structure with
a narrow domain wall (DW) at the SF interface. The width of the domain wall
is assumed to be larger than the Fermi wave length, but smaller than other
characteristic lengths (for example, the ''magnetic'' length). The
transmission coefficient is supposed to be small so that we deal with a weak
proximity effect. Solving the linearized Eilenberger equation, we find
analytical expressions for quasiclassical Green's functions. These functions
describe the short-range (SR) condensate components, singlet and triplet
with zero projection of the total spin on the quantization $z$-axis, induced
in F due to the proximity effect as well as long-range odd triplet component
(LRTC) with a nonzero projection of the total spin of Cooper pairs on the $z$%
-axis. The amplitude of the LRTC essentially depends on the product $h\tau $
and increases with increasing the exchange energy $h$ ($\tau $ is the
elastic scattering time). We calculate the Josephson current in SFS junction
with a thickness of the F layer much greater than the penetration length of
the SR components. The Josephson critical current caused by the LRTC may be
both positive and negative depending on chirality of the magnetic structure
in F.

The density of states (DOS) in a diffusive SF bilayer is also analyzed. It
is shown that the contributions of the SR and LR components to the DOS in F
have a different dependence on the thickness $d$ of the F layer
(nonmonotonous and monotonous).
\end{abstract}

\maketitle

\bigskip

\section{Introduction}

For a long time the mechanism of superconductivity of Bardeen, Cooper and
Schrieffer \cite{BCS,Schrieffer} based on the assumption of\textbf{\ }s-wave
singlet pairing\textbf{\ }remained sufficient for explanation of\textbf{\ }%
properties of almost all existing superconductors. However\textbf{, }the
situation has changed in the last two decades. It has been established that
in high $T_{c}$ superconductors the d-wave singlet pairing is responsible
for the superconductivity \cite{Kirtley}. A triplet p-wave pairing was
suggested to explain properties of materials with heavy fermions \cite%
{Mineev}. Recent intensive studies of the strontium ruthenate (Sr$_{2}$RuO$%
_{4}$) showed that superconductivity in this compound was also\textbf{\ }due
to the triplet p-wave pairing mechanism \cite{Maeno,Eremin}. The p-wave
pairing was considered to be an essential ingredient for forming the triplet
Cooper pairs because, in contrast to the conventional singlet pairing, it
allowed to satisfy the Pauli principle.\textbf{\ }

A more exotic type of triplet pairing was proposed by Berezinskii \cite%
{Berez} as a possible mechanism of superfluidity of $He^{3}$. According to
this suggestion the triplet pairing might have a singlet space symmetry.%
\textbf{\ }The wave function of the Cooper pairs, $f_{\uparrow \uparrow
}(t,t^{\prime })$ $\propto \langle \psi _{\uparrow }(t)\psi _{\uparrow
}(t^{\prime })\rangle $, suggested by Berezinskii was symmetric in both%
\textbf{\ }momentum and spin space. At the same time\textbf{, }in order to
fulfill the Pauli principle, the function $f_{\uparrow \uparrow }(t,t)$
taken at equal times must be zero. The only possibility to satisfy all these
requirements is to assume that the wave function in the Matsubara
representation $f_{\uparrow \uparrow }(\omega )$ should be an odd function
of $\omega $ so that $f_{\uparrow \uparrow }(t,t)\propto \sum_{\omega
}f_{\uparrow \uparrow }(\omega )=0$. This is exactly what was suggested by
Berezinskii and one may call such a state odd triplet superconductivity
(superfluidity)\textbf{. }

Unfortunately, this type of pairing was not more than a hypothesis and no
microscopic model leading to the odd triplet superconductivity was suggested
in Ref. \cite{Berez}. Moreover,\textbf{\ }it turned out later that in
superfluid $He^{3}$ another type of pairing was responsible for the
superfluidity \cite{Leggett,Wolfle} and the scenario for the odd triplet
superconductivity remained justified neither theoretically nor
experimentally.\textbf{\ }

It was discovered only recently \cite{BVE01} that the odd triplet
superconductivity could be realized in a simple system consisting of an
ordinary BCS superconductor (S) and ferromagnet (F) with a nonhomogeneous
magnetization $M$ (for details, see also a review \cite{BVErmp} and
references therein). In case of an SF system (for example, an SF bilayer)
with a homogeneous magnetization in F, two types of the condensate arise in
the system - a singlet component with a condensate wave function $f_{3}$ and
a triplet component $f_{0}$ with the zero projection of the total spin of
the Cooper pair on the quantization axis, $S_{z}=0$ \cite{BuzdinRMP,BVErmp}.
Both the\textbf{\ }components decay in F over a short length
\begin{equation}
\xi _{h}=\sqrt{D/h}  \label{e1}
\end{equation}%
in the diffusive limit ($h\tau <<1$) and over the mean free path $l=v\tau $
in the limit $h\tau >>1$, where $D=vl/3$ is the diffusion coefficient and $h$
is the exchange energy. Since the exchange energy $h$ is much larger than
the temperature $T$, the decay length $\xi _{h}$ is much shorter than the
depth of the condensate penetration into a nonmagnetic metal N in case of an
SN system.

In Ref. \cite{BVE01} a diffusive SF system with a domain wall (DW) at the SF
interface was considered. It was shown that in this case not only the
singlet and triplet $S_{z}=0$ components but also a triplet $|S_{z}|=1$
component arises in the system. The triplet component with a nonzero
projection of the total spin $S_{z}$ penetrates the ferromagnet over a
length $\xi _{\omega }$ that does not depend on the exchange energy $h$ and
equals
\begin{equation}
\xi _{\omega }=\sqrt{D/2\omega }  \label{e2}
\end{equation}%
in the diffusive limit, where $\omega =\pi T(2n+1)$ is the Matsubara
frequency. This triplet component was called a long-range triplet component
(LRTC).

In order to carry out calculation of physical quantities explicitly\textbf{,
}it was assumed in Ref. \cite{BVE01} that the width of the domain wall $w$
was much larger than the mean free path $l$ and that\textbf{\ }the proximity
effect was weak due to a non-ideal SF interface (the reflection coefficient
at the interface $R$ is close to 1). The magnetization vector $\mathbf{M}$
in the domain wall was assumed to rotate linearly with the coordinate $x$
normal the the SF interface so that the angle $\alpha $ between $\mathbf{M}$
and $z$-axis was equal to $\alpha (x)=Qx$ in the interval $\{0,w\}$ and $%
\alpha =Qw$ at $x>w$. In this case the condensate functions in F could be%
\textbf{\ }found exactly from the linearized Usadel equation.

As a result,\textbf{\ }the amplitude $f_{1}$ of the odd triplet $|S_{z}|=1$
component has been obtained in Ref. \cite{BVE01}explicitly. Using the known
value of the function $f_{1}$, the conductance variation $\delta G$ of the
ferromagnet as a function of temperature $T$ has been determined. It turned
out that, in contrast to the SN system where the conductance variation has a
maximum at some temperature (a reentrant behavior) \cite%
{AVZ,GubMar,Naz,LamVol,Pann}, the function $\delta G(T)$ in the SF system
decreased monotonously with increasing the temperature.

In subsequent works \cite{Kadig,Eschrig,Tanaka,Zaikin} the idea about the
generation of the odd triplet condensate with the non-zero projection and
long range penetration into the ferromagnets was discussed using somewhat
different models and approaches.\textbf{\ }

The authors of Ref. \cite{Kadig} also considered a model with a DW at the SF
interface but, in contrast to Ref. \cite{BVE01},\textbf{\ }assumed that the
length of the DW was shorter than the mean free path. Although nothing was
said in the paper about the type of the SF interface, they considered
apparently the limit of the ideal SF interface (the transmission coefficient
did not enter the equations presented in that paper). It was assumed that in
the region of the domain wall the superconducting condensate had to be
described by an Eilenberger equation. At distances exceeding the mean free
path one should have the Usadel equation and the Eilenberger equation might
be used to derive a boundary condition for the Usadel equation.\textbf{\ }

Without presenting a solution of the Eilenberger equation the authors of
Ref. \cite{Kadig} displayed in a simple form an effective boundary condition
for the linearized Usadel equation. This boundary condition introduced the
triplet condensate as a solution of the Usadel equation and the latter was
used to determine the contribution to the conductivity due to the triplet
condensate penetrating the ferromagnet over long distances.

Another approach to finding the odd triplet component was suggested in Refs.%
\cite{Eschrig,Eschrig08,Tanaka}. In that approach the properties of the SF
interface are characterized by a scattering matrix the elements of which may
be considered as phenomenological parameters. In this approach one does not
need knowing the detailed structure of the SF interface and can proceed
calculating physical quantities using these parameters.\textbf{\ }The
amplitude of the condensate wave functions has been determined in these
papers numerically. Analytical results were obtained in the framework of
this approach in a recent paper \cite{Zaikin}, where a ballistic SFS system
was considered.

However, from the physical point of view this approach is equivalent to
introducing a thin domain wall. If one wants to know details of how the
triplet component is generated one has to solve again the microscopic
Eilenberger equation for a certain configuration of the magnetic moment,
which is similar to considering the model of Ref. \cite{Kadig}.

In this paper, we reconsider the problem of the generation of the odd
triplet component by a thin domain wall located at the SF interface. We
assume that the size the domain wall exceeds the interatomic distances,
which allows us to use the quasiclassical Eilenberger equation \cite%
{Eilenberger,LO}.\textbf{\ }Below, we solve the Eilenberger equation
assuming a weak proximity effect and show that some effective boundary
condition for the Usadel equation can indeed be written. However, the
results we obtain disagree with those of Ref. \cite{Kadig}. It turns out
that, in contrast to the formula of Ref. \cite{Kadig}, the effective
boundary condition\textbf{\ }for the Usadel equation crucially depends on
the relation between the exchange energy $h$, elastic scattering rate $\tau
^{-1}$ and other parameters. Moreover, the absence of the transmission
coefficient $T(\mu )$ in formulas of Ref.\cite{Kadig} makes us to suppose
that the SF interface was assumed to be ideal. However, in this case one has
to solve non-linearized Eilenberger equation and we believe that this can be
done only numerically. Therefore we suspect that the form of the boundary
condition presented in Ref. \cite{Kadig} is not well justified.\textbf{\ }

By now, several attempts to observe this new type of the condensate - odd
triplet component - have been undertaken. In a recent work \cite{Keiser} the
dc Josephson effect has been measured in an SFS Josephson junction
consisting of two superconductors (Nb) and \ the ferromagnet CrO$_{2}$ where
free electrons have only one direction of spins. The Josephson critical
current has been observed in junctions with a separation between S
electrodes of about $1\mu m$. Obviously, the Josephson coupling between the
superconductors may only be due to the LRTC. In Ref.\cite{Sosnin} a
conductance variation $\delta G$ was measured in an Al/Ho system below the
critical temperature of Al. The order of magnitude of the observed change of
the conductance can really be explained in terms of the LRTC. In this
ferromagnet, a magnetic inhomogeneity is natural because Ho is a helicoidal
ferromagnet such that the magnetization vector rotates in space forming a
spiral with the period $\approx 60A$.

Already earlier experiments on SF structures have also brought an evidence
in favor of the existence of a condensate penetrating into the ferromagnet
over a long distance \cite{Lawrence,Petrashov99,Giraud,Chandrasekhar01}. It
is also worth mentioning that in recent experiments on SFS junctions \cite%
{Ryaz,Kontos,Blum,Bauer}, where the sign-reversal of the critical current ($%
\pi -$state) has been detected, the magnetization was not homogeneous, and
therefore the triplet component had also to exist and contribute to the
critical current. The problem of the triplet component in multilayered SFS
junctions \cite{BVE03,Buzdin07,NazBr,Sudbo,Champel08} and in junctions with
Neel's domain walls \cite{Fominov} was studied in recent theoretical papers.
It was shown in particular that the LRTC may also lead to a non-monotonous
dependence of the critical Josephson current in SFS junctions \cite%
{Anishchanka06,Champel08}. In order to succeed in searching the LRTC in SF
structures it is very important to use materials that might give a large
amplitude of the LRTC. Therefore\textbf{, }it is desirable to have
analytical formulas for the amplitude of the LRTC in a wide range of
parameters characterizing the system.\textbf{\ }Calculation of these
amplitudes is the ultimate goal of the present work.

The paper is organized as follows. In Sec. II we formulate the model and
solve the linearized Eilenberger equation. Expressions for short and
long-range condensate components (SR and LR) induced in F are also presented
there. In Sec. III spatial dependencies of the SR and LR components are
found for a weak ($h\tau <<1$) and strong ($h\tau >>1$) ferromagnets. The
Josephson current in a long SFS junction originating from the LRTC is
calculated in Sec. IV. In Sec. V we analyze a diffusive SF bilayer with a DW
the width of which exceeds the mean free path but is shorter than the
\textquotedblleft magnetic\textquotedblright\ length $\xi _{h}$. The
influence of the spin-dependent scattering on the LRTC is also analyzed in
this section. In Sec.\ VI we consider a diffusive SF bilayer. We calculate
the contributions to the density of states (DOS) due to the SR and LR
components and discuss a possible reason for an anomalous behavior of the
DOS observed in a recent experiment \cite{Beasley}. In Conclusions we
discuss the results obtained.

\bigskip

\section{Model and basic equations}

We consider an SF structure with a thin domain wall at the SF interface (see
Fig.1). The thickness of the DW, $w,$ is supposed to be larger than the
Fermi wave length but smaller than all other characteristic lengths (the
mean free path $l=v\tau $, the \textquotedblleft exchange
length\textquotedblright\ $v/h$ etc). Outside the interval $\{0,w\}$ the
magnetization vector $\mathbf{M}$ in F is parallel to the $z$-axis but
inside the domain wall it has a projection on the $y$-axis $M\sin \alpha (x)$%
. The average $<\sin \alpha (x)>_{w}\equiv \frac{1}{w}\int_{0}^{w}dx\sin
\alpha (x)\neq 0$ is assumed to differ from zero. The transmission of
electrons through the SF interface is supposed to be small so that we deal
with a weak proximity effect. This assumption seems to correspond to
experiments because even in the absence of a potential barrier at the SF
interface the reflection of electrons at the interface is strong due to a
considerable mismatch of the Fermi surfaces in the superconductor and
ferromagnet.

\begin{figure}[tbp]
\begin{center}
\includegraphics[width=25cm, height=7cm]{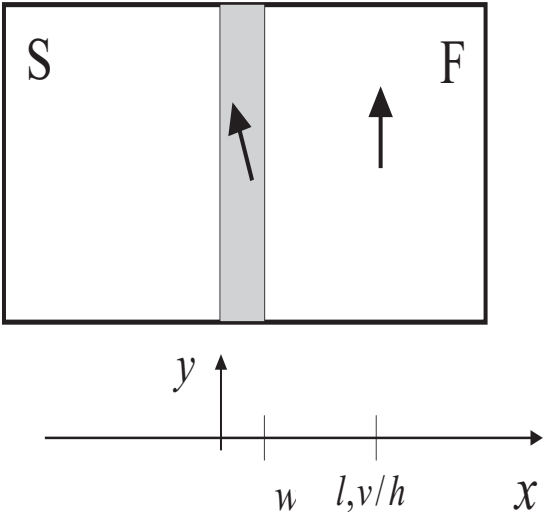}
\end{center}
\par
{\large {} }
\caption{Schematic picture of a SF bilayer with a domain wall (the shadowed
stripe) at the SF interface. The width of the DW is $w$; $v/h,l$ denote the
"magnetic" length and the mean free path. }
\end{figure}

Our calculations are based on the Eilenberger equation for quasiclassical
Green's functions \cite{Eilenberger,LO,Kopnin}. In the limit of the weak
proximity effect considered here\textbf{, }these functions in S and F
deviate weakly from their values in the absence of the contact between S\
and F. We are interested here in the condensate wave functions induced in F.
Due to the presence of the exchange field acting on spins of free electrons,
the system should be described by quasiclassical Green's functions $\check{g}
$ that are $4\times 4$ matrices in the particle-hole and spin space.

The Eilenberger equation in the ferromagnet F has the form \cite%
{Eilenberger,LO,Kopnin,BVErmp}

\begin{equation}
iv\nabla \check{g}+\omega \lbrack \hat{\tau}_{3}\otimes \hat{\sigma}_{0},%
\check{g}]+i[\mathbf{h(}x\mathbf{)S},\check{g}]+(i/2\tau )\left[ \langle
\check{g}\rangle ,\check{g}\right] =0\;.  \label{Eilenberger0}
\end{equation}%
where $\mathbf{h(}x\mathbf{)=}h(0,\sin \alpha (x),\cos \alpha (x))$ is the
vector of the exchange field, $\omega =\pi T(2n+1)$ is the Matsubara
frequency, $v$ is the Fermi velocity, $\mathbf{S}=(\hat{\sigma}_{1},\hat{%
\sigma}_{2},\hat{\tau}_{3}\otimes \hat{\sigma}_{3}),$ $\hat{\sigma}_{k}$ are
the Pauli matrices and $\hat{\sigma}_{0}$ is the unit matrix. The square and
angle brackets mean the commutator and averaging over angles, respectively,
and $\tau $ is the elastic scattering time. The matrices $\hat{\tau}_{k}$
and $\hat{\sigma}_{k}$ operate in the particle-hole and spin space. As has
been assumed previously, outside of the DW ($x>w$) the angle $\alpha $ is
just zero, $\alpha (x)=0$. At the same time, the concrete spatial dependence
of the angle $\alpha $ inside the DW is not essential. It is important only
that
\[
<\sin \alpha (x)>_{w}\equiv \frac{1}{w}\int_{0}^{w}dx\sin \alpha (x)\neq 0
\]%
Eq. (\ref{Eilenberger0}) is complemented by a boundary condition \cite%
{Zaitsev}

\begin{equation}
\check{a}\equiv (\check{g}(\mu )-\check{g}(-\mu ))/2=\mathit{sgn}\mu \cdot
(T(\mu )/4)[\check{g},\check{g}_{S}]  \label{BCo}
\end{equation}%
where $\check{a}$ is the part of the quasiclassical Green function $\check{g}
$ antisymmetric in the momentum space, $\mu =p_{x}/p,$ $\check{g}=\check{g}%
_{F}$ is the Green's function in F, and $T(\mu )$ is the coefficient of
transmission of electrons through the SF interface which is supposed to be
small.

Due to the weakness of the proximity effect, it is convenient to represent
the quasiclassical Green function $\check{g}$ in the form

\begin{equation}
\check{g}=\mathit{sgn}\omega \cdot \hat{\tau}_{3}\otimes \hat{\sigma}_{0}+%
\check{f}\;.  \label{gNorm}
\end{equation}%
where the first term is the Green's function of the ferromagnet in the
absence of the superconducting condensate and the second term is the
condensate function we are interested in. Substituting $\check{g},$ Eq.(\ref%
{gNorm}), into Eq. (\ref{Eilenberger0}) and linearizing the Eilenberger
equation with respect to $\check{f}$, we come to the equation

\begin{equation}
\mathit{sgn}\omega \cdot \mu \hat{\tau}_{3}\mathbf{\otimes }\partial \check{f%
}\;/\partial \tilde{x}+(1+2|\omega |\tau )\check{f}\;-i(h_{\omega }\tau
)\cos \alpha (x)[\hat{\sigma}_{3},\check{f}]_{+}=\langle \check{f}\rangle
\;+i(h_{\omega }\tau )\sin \alpha (x)\hat{\tau}_{3}\mathbf{\otimes }[\hat{%
\sigma}_{2},\check{f}].  \label{EilenLin}
\end{equation}%
where $\tilde{x}=x/l$ is the dimensionless coordinate $h_{\omega }=h\cdot
\mathit{sgn}\omega ,$ the brackets\textbf{\ }$[..,..]_{+},[..,..]$ stand for
the anticommutator and commutator, respectively, and the angle brackets mean
the averaging over angles. When writing Eq. (\ref{EilenLin}), we used the
fact that the condensate matrix function $\check{f}$ is off-diagonal in the
particle-hole space and therefore anticommutes with the matrix $\hat{\tau}%
_{3}$. We neglect here the spin-dependent scattering caused by fluctuations
of magnetic moments in space and spin-orbit interaction. The influence of
this scattering will be discussed in section 5.

For small values of the condensate function $\check{f}$ the boundary
condition, Eq. (\ref{BCo}), can also be linearized and written as

\begin{equation}
\check{a}=sgn\mu \cdot sgn\omega \cdot (T(\mu )/2)\hat{\tau}_{3}\check{f}_{S}
\label{BCoLin}
\end{equation}
where $\check{f}_{S}=\hat{\sigma}_{3}\mathbf{\otimes }\hat{\tau}_{2}f_{S}$
is the condensate matrix function in S in the absence of the proximity
effect, $f_{S}=\Delta /\sqrt{\omega ^{2}+\Delta ^{2}}$.

So, we\textbf{\ }have to solve Eq.(\ref{EilenLin}) with the boundary
condition, Eq. (\ref{BCoLin}). To find the solution we represent the matrix $%
\check{f}\;$as the sum of symmetric $\check{s}$ and antisymmetric $\check{a}$
in the\textbf{\ }momentum space parts

\begin{equation}
\check{f}=\check{s}+\check{a}  \label{s+a}
\end{equation}

Substituting the representation for the matrices $\check{s}$\ and $\check{a}$%
, Eq. (\ref{s+a}), into Eq.(\ref{EilenLin}),\textbf{\ }we come to the
following equations

\begin{eqnarray}
\mathit{sgn}\omega \cdot \mu \hat{\tau}_{3}\mathbf{\otimes }\partial
\check{s}\;/\partial \tilde{x}+\kappa _{\omega }\check{a}\;-i(h_{\omega
}\tau )\cos \alpha (x)[\hat{\sigma}_{3},\check{a}]_{+} &=&i\hat{\tau}%
_{3}(h_{\omega }\tau )\sin \alpha (x)[\hat{\sigma}_{2},\check{a}],
\label{EqS} \\
\mathit{sgn}\omega \cdot \mu \hat{\tau}_{3}\mathbf{\otimes }\partial \check{a%
}/\partial \tilde{x}+\kappa _{\omega }\check{s}\;-i(h_{\omega }\tau )\cos
\alpha (x)[\hat{\sigma}_{3},\check{s}]_{+} &=&\langle \check{s}\rangle +i%
\hat{\tau}_{3}(h_{\omega }\tau )\sin \alpha (x)[\hat{\sigma}_{2},\check{s}],
\label{EqA}
\end{eqnarray}
where $\kappa _{\omega }=1+2|\omega |\tau $.

If we neglected the right-hand side, the solution of Eqs.(\ref{EqS},\ref{EqA}%
) with the boundary condition (\ref{BCoLin}) would contain only the singlet
and $S_{z}=0$ triplet components. The presence of the right-hand side of
Eqs. (\ref{EqS}, \ref{EqA}) results in the appearance of the LRTC. If the
domain width $w$ is small in comparison with\textbf{\ }the other
characteristic lengths of the problem, $v/h$ and $l$, all the functions vary
slowly over this distance. Therefore, we can integrate Eq.(\ref{EqA}) over
the interval $\{0,w\}$ and obtain an effective boundary condition for the
matrix $\check{a}$

\begin{equation}
\mu \check{a}|_{x=0}=\mathit{sgn}\omega \cdot b_{\mu }\hat{\tau}_{3}\mathbf{%
\otimes }\hat{\sigma}_{3}\hat{f}_{S}+i\mathit{H}[\hat{\sigma}_{2},\check{s}%
(0)]  \label{BCoEff}
\end{equation}
where $b_{\mu }=(T(\mu )|\mu |/2),$ \ $\hat{f}_{S}=f_{S}\hat{\tau}_{2}$, and
$\mathit{H}=(h\tau )(w/l)\langle \sin \alpha (x)\rangle _{w}\equiv (h\tau )(%
\bar{w}/l),$ $\bar{w}=w\langle \sin \alpha (x)\rangle _{w}.$ For example, in
the case of DW with a linearly varying magnetization we obtain $\bar{w}%
=w(2/\pi )\approx 0.64w$.

Now the problem is reduced to solving Eqs. (\ref{EqS},\ref{EqA}) outside the
domain wall ($\alpha =0$) with the boundary condition (\ref{BCoEff}). We
will see that the symmetric part, $\check{s}$, has the following structure
in the spin space

\begin{equation}
\check{s}=(\hat{s}_{3}\hat{\sigma}_{3}+\hat{s}_{0}\hat{\sigma}_{0})+\hat{s}%
_{1}\hat{\sigma}_{1}  \label{StrS}
\end{equation}
with $\hat{s}_{3}=s_{3}\hat{\tau}_{2},$ $\hat{s}_{0}=s_{0}\hat{\tau}_{2}$,
and $\hat{s}_{1}=s_{1}\hat{\tau}_{1}$.

First, we consider elements of the $\check{s}$ matrix diagonal in the spin
space. From Eq. (\ref{EqS}) we find for $a_{\pm }\equiv a_{11(22)}$

\begin{equation}
a_{\pm }\cdot \kappa _{h\pm }=\mathit{sgn}\omega \cdot |\mu |\hat{\tau}_{3}%
\mathbf{\cdot }\partial \hat{s}_{\pm }/\partial \tilde{x}  \label{aPM}
\end{equation}%
where $\kappa _{h\pm }=1+2|\omega |\tau \mp 2ih_{\omega }\tau $.

Using Eq. (\ref{aPM}) the effective boundary condition, Eq. (\ref{BCoEff}),
can be written as

\begin{equation}
-\mu ^{2}\partial \hat{s}_{\pm }/\partial \tilde{x}=\pm \kappa _{h\pm
}[b_{\mu }\hat{f}_{S}+2\mathit{H}_{\omega }\cdot \hat{\tau}_{3}\cdot \hat{s}%
_{1}(0)]  \label{BCoEffS}
\end{equation}
where $\mathit{H}_{\omega }=\mathit{H}\cdot \mathit{sgn}\omega $.

Substituting $a_{\pm }$ from Eq. (\ref{aPM}) into Eq. (\ref{EqA}), one can
write an equation for $\hat{s}_{\pm }$ in the form

\begin{equation}
-\mu ^{2}\partial ^{2}\hat{s}_{\pm }/\partial \tilde{x}^{2}+\kappa _{h\pm
}^{2}\hat{s}_{\pm }=\kappa _{h\pm }\langle \hat{s}_{\pm }\rangle \;\pm
2\delta (\tilde{x})\kappa _{h\pm }[b_{\mu }\hat{f}_{S}+2\mathit{H}_{\omega }%
\hat{\tau}_{3}\cdot \hat{s}_{1}(0)]  \label{EqSpm}
\end{equation}
The boundary condition, Eq. (\ref{BCoEffS}), is taken into account with the
help of the last term in the right-hand side of Eq. (\ref{EqSpm}) and a
formal symmetric continuation of the solution to the interval $\{-\infty
,0\} $. Performing the same procedure we can obtain an equation for $\hat{s}%
_{1}$

\begin{equation}
-\mu ^{2}\partial ^{2}\hat{s}_{1}/\partial \tilde{x}^{2}+\kappa _{\omega
}^{2}\hat{s}_{1}=\kappa _{\omega }\langle \hat{s}_{1}\rangle \;-4\delta (%
\tilde{x})\kappa _{\omega }\mathit{H}_{\omega }\hat{\tau}_{3}\cdot \hat{s}%
_{3}(0)  \label{EqS1}
\end{equation}

Eq. (\ref{EqS1}) can easily be\textbf{\ }solved in the same way as it was
done for the case of a homogeneous magnetization \cite{BVECrCur}. For the
Fourier transforms $\hat{S}_{\pm }(k)$, $\hat{S}_{1}(k)$ of the functions%
\textbf{\ }$\hat{s}_{\pm }$, $\hat{s}_{1}$ we obtain

\begin{equation}
\hat{S}_{\pm }(k)=\pm 2\frac{\kappa _{h\pm }}{M_{h\pm }(k,\mu )}{\LARGE \{}%
\frac{\kappa _{h\pm }}{1-\kappa _{h\pm }\langle M_{h\pm }^{-1}(k,\mu
)\rangle }\langle \frac{b_{\mu }\hat{f}_{S}+\mathit{H}_{\omega }\hat{\tau}%
_{3}\cdot \hat{s}_{1}(0)}{M_{h\pm }(k,\mu )}\rangle \;+b_{\mu }\hat{f}_{S}+%
\mathit{H}_{\omega }\hat{\tau}_{3}\cdot \hat{s}_{1}(0){\LARGE \}}
\label{sPMk}
\end{equation}

\bigskip
\begin{equation}
\hat{S}_{1}(k)=-4\frac{\mathit{H}_{\omega }\kappa _{\omega }}{M_{\omega
}(k,\mu )}\hat{\tau}_{3}{\LARGE \{}\frac{\kappa _{\omega }}{1-\kappa
_{\omega }\langle M_{\omega }^{-1}(k,\mu )\rangle }\langle \frac{\hat{s}%
_{3}(0)}{M_{\omega }(k,\mu )}\rangle \;+\hat{s}_{3}(0){\LARGE \}}
\label{s1k}
\end{equation}%
where $M_{h\pm }(k,\mu )=(k\mu )^{2}+\kappa _{h\pm }^{2},$ $M_{\omega
}(k,\mu )=(k\mu )^{2}+\kappa _{\omega }^{2},$ $\mathit{H}_{\omega }=\mathit{%
sgn}\omega (h\tau )(w/l)\langle \sin \alpha \rangle _{w}\equiv \mathit{sgn}%
\omega (h\tau )(\bar{w}/l),$ $\kappa _{h\pm }=1+2|\omega |\tau \mp
ih_{\omega }\tau ,$ $\kappa _{\omega }=1+2|\omega |\tau $.

One can see from Eq.(\ref{s1k}) that the characteristic length of the decay
of the LRTC, $s_{1}(x)$, does not depend on the exchange energy $h$ \cite%
{BVE01}. We will see that the spin dependent scattering makes the
characteristic decay length shorter. Account for this scattering changes the
quantity $\kappa _{\omega }$ as $\kappa _{\omega }\Rightarrow \kappa
_{\omega }=1+2|\omega |\tau +\lambda _{\perp }+(4/9)\lambda _{so}$ (see
Sec.V). As follows from Eqs.(\ref{sPMk}-\ref{s1k}), the SR components, $%
\hat{S}_{\pm }$, arise in the case of a homogenous magnetization when $%
\mathit{H}_{\omega }=0$. The LRTC appears only in the presence of a
nonhomogeneous magnetization, for example in the presence of a DW when $%
\mathit{H}_{\omega }\sim (h\tau )(\bar{w}/l)\neq 0$.

Eqs. (\ref{sPMk}, \ref{s1k}) are the main results of the paper. They
determine the spatial dependence of the short, $\hat{s}_{\pm }$, and long, $%
\hat{s}_{1}$, range amplitudes of the condensate. Note that the amplitudes
of the singlet, $\hat{S}_{3}(k),$ and short-range triplet, $\hat{S}_{0}(k)$,
components are expressed through $\hat{S}_{\pm }(k)$ in a simple way

\begin{equation}
\hat{S}_{0,3}(k)=(\hat{S}_{+}(k)\pm \hat{S}_{-}(k))/2  \label{S03}
\end{equation}

Although Eqs. (\ref{sPMk}, \ref{s1k}) completely determine the solutions of
Eqs. (\ref{EqSpm}, \ref{EqS1}), the explicit form of the solutions is still
to be obtained from the inverse Fourier transform. Unfortunately, the latter
can be presented in an analytical form only in some limiting cases. In the
next section we will analyze the spatial dependence of the amplitudes $%
\hat{s}_{0,1,3}(x).$

\section{Spatial dependence of the condensate wave functions}

Using Eqs. (\ref{sPMk}, \ref{s1k}) one can obtain the spatial dependence of
the amplitudes $\hat{s}_{0,1,3}(x)$ describing the penetration of the odd
triplet condensate into the ferromagnet. The corresponding expressions are
to be found by calculating the inverse Fourier transform of Eqs. (\ref{sPMk}%
, \ref{s1k}).

The form of the expressions for $\hat{S}_{\pm }(k)$, $\hat{S}_{1}(k)$
indicates\textbf{\ }that the spatial dependence of the amplitudes $\hat{s}%
_{0,1,3}(x)$ is determined by zeros of the functions $M_{\omega }(k,\mu )$
and $M_{h\pm }(k,\mu )$ as well as of the functions $(1-\kappa _{\omega
}\langle M_{\omega }^{-1}(k,\mu )\rangle )$ and $(1-\kappa _{h\pm }\langle
M_{h\pm }^{-1}(k,\mu )\rangle )$. Although the decay length of the
amplitudes $\hat{s}_{0,3}(k)$ depends on the exchange energy $h$, the decay
length of the amplitude $\hat{s}_{1}(k)$ does not.

The LRTC in the Fourier representation, $\hat{S}_{1}(k),$ is expressed
through the short-range singlet component $\hat{s}_{3}(0)$ at $x=0$, and in
its turn, the matrix $\hat{S}_{\pm }(k)$ depends on the amplitude of the
singlet component in S, $\hat{f}_{S},$ and on the LRTC $\hat{s}_{1}(0)$ at $%
x=0.$ We suppose that the influence of the LRTC on the short-range amplitude
$\hat{S}_{\pm }(k)$ is weak, that is, the condition

\begin{equation}
\mathit{H}|s_{1}(0)|<<b_{\mu }f_{S}  \label{Cond1}
\end{equation}
is satisfied.

The matrices $\hat{S}_{\pm }(k)$ and $\hat{S}_{1}(k)$ may be found from Eqs.(%
\ref{sPMk}) and (\ref{s1k}), respectively. Then, performing the integration
over $k,$ one can find $s_{0,3}(0)=\int (dk/2\pi )s_{0,3}(k)$ and $s_{1}(0)$%
. However, the expressions for these matrices are too cumbersome even if the
condition (\ref{Cond1}) is fulfilled. One can further simplify these
expressions considering the limits of large and small products $h\tau ,$
i.e., considering the case of a strong or weak ferromagnet. We also will
assume that the condition

\begin{equation}
T\tau <<1  \label{Cond2}
\end{equation}%
is fulfilled because it corresponds to experimental situations.
Corresponding formulas can easily be\textbf{\ }obtained also in the opposite
limit.

First, we consider the limit of a weak ferromagnet when the inequality

\textit{a) }$\left\{ h,T\right\} <<\tau ^{-1}$ is fulfilled\textbf{\ }(the
diffusive case or the case of a weak ferromagnet).

In this case, the main contribution comes from small $k$ ($k<<1$) \cite%
{BVECrCur} and we obtain: $1-\kappa _{h\pm }\langle M_{h\pm }^{-1}(k,\mu
)\rangle \approx (k^{2}+K_{h}^{2})/3$ and $1-\kappa _{\omega }\langle
M_{\omega }^{-1}(k,\mu )\rangle \approx (k^{2}+K_{\omega }^{2})/3$, where $%
K_{h\pm }^{2}=6(|\omega |\mp ih_{\omega })\tau =l\sqrt{(2|\omega |\mp
ih_{\omega })/D}$ and $K_{\omega }^{2}=3(\kappa _{\omega }-1).$ Calculating
the residue of the pole at $k=iK_{h}$ in Eq.(\ref{sPMk}), we find for the
amplitudes of the SR components
\begin{equation}
\hat{s}_{\pm }(x)=\pm 3(b_{\mu }/K_{h\pm })\hat{f}_{S}\exp (-x/\xi _{h\pm
})\;  \label{DifSpm}
\end{equation}%
where $\xi _{h\pm }=\sqrt{D/[2(|\omega |\mp ih_{\omega })]}$ is the
characteristic length over which the short range components (singlet and
triplet ones with zero projection $S_{z}$ on the $z$ axis) penetrate the
ferromagnet.

The LRTC can be found from Eq. (\ref{s1k}) calculating the residue and we
obtain

\begin{equation}
\hat{s}_{1}(x)=-18\mathit{H}_{\omega }\langle b_{\mu }\rangle (\xi _{LR}/l)Re%
\frac{1}{K_{h}}(\hat{\tau}_{3}\cdot \hat{f}_{S})\exp (-x/\xi _{LR})\;
\label{DifS1}
\end{equation}
with $\xi _{LR}=\sqrt{D/(2|\omega |\tau +\lambda _{\perp }+(4/9)\lambda
_{so})}$ and $K_{h}=\sqrt{6(|\omega |-ih_{\omega })\tau }$.

Eqs. (\ref{DifSpm}, \ref{DifS1}) describe \ the spatial dependence of the
short- and long-range components of the condensate. The SR component $\hat{s}%
_{\pm }(x)\sim \hat{\tau}_{2}\exp (-x/\xi _{h\pm })$ decays over a short
length, $\xi _{h\pm },$ and experiences oscillations \cite%
{BuzdinRMP,BVErmp,GolubovRMP}. The LRTC $\hat{s}_{1}(x)\sim \hat{\tau}%
_{1}\exp (-x/\xi _{\omega })$ decays without oscillations over a long
distance $\xi _{LR}$ \cite{BVE01,BVErmp,BuzdinRMP}. The amplitude of the
singlet component $\hat{s}_{3}(0)=s_{3}\hat{\tau}_{2}$ at $x=0$ equals

\begin{equation}
\hat{s}_{3}(0)=3b_{\mu }Re\frac{1}{K_{h}}\hat{f}_{S}.\;
\end{equation}%
\textbf{\ }Thus, the ratio of the LRTC $\hat{s}_{1}(0)=s_{1}\hat{\tau}_{1}$
to the singlet component $s_{3}(0)$ takes the form

\begin{equation}
r=|\frac{s_{1}(0)}{s_{3}(0)}|=2\frac{\xi _{LR}}{\xi _{h}^{2}}\;\bar{w}
\label{RatioDiff}
\end{equation}%
This ratio may be both larger and smaller than $1$. The amplitude of the
LRTC increases\textbf{\ }with increasing the exchange energy $h$.

The condition (\ref{Cond1}) can be rewritten in this limit as

\begin{equation}
(\frac{\bar{w}}{l})<<\frac{1}{3\sqrt{6}}\frac{\xi _{LR}}{(h\tau )^{3/2}l}
\label{CondDiff}
\end{equation}

If the spin-coupling constant $\lambda _{so}$ is larger than the product $%
|\omega |\tau ,$ this inequality can be written as $h\tau <<(1/18)\lambda
_{so}$.

Now we consider the limit of the large exchange energy $h$:

\textit{b) }$T<<\tau ^{-1}<<h$ (the case of a strong ferromagnet).

In this case the quantity $\kappa _{h\pm }\langle M_{h\pm }^{-1}(k,\mu
)\rangle $ is small because $|\kappa _{h\pm }|>>1$. Therefore, the main
contribution to $\hat{s}_{\pm }(x)$ is due to the second term in the figure
brackets in Eqs.(\ref{sPMk}) and one has to calculate the residue of the
pole of the functions $\left( M_{h\pm }\left( k,\omega \right) \right) ^{-1}$%
. The formula for $\hat{s}_{1}(x)$ is obtained as before.

As a result\textbf{, }we find

\begin{equation}
\hat{s}_{\pm }(x)=\pm (b_{\mu }/\mu )\hat{f}_{S}\exp (-\kappa _{h\pm }x/l)\;
\end{equation}%
and

\begin{equation}
\hat{s}_{1}(x)=-6\langle b_{\mu }/\mu \rangle \mathit{H}_{\omega }(\xi
_{LR}/l)(\hat{\tau}_{3}\cdot \hat{f}_{S})\exp (-x/\xi _{LR})\;
\end{equation}

The SR components $\hat{s}_{\pm }(x)$ oscillate with the period $\pi v/h$
and decrease in the ferromagnet over the mean free path $l$ as has been
obtained earlier in this limit \cite{BVECrCur,BB,BVE02}. The LRTC decreases
in a monotonic way over the length $\xi _{LR}$. The ratio of the amplitude
of the LRTC to the short range singlet component at $x=0$ is equal to

\begin{equation}
r=6\;(\frac{\bar{w}}{v/h})\frac{\xi _{LR}}{l}  \label{rLargeH}
\end{equation}%
and\textbf{\ }the condition (\ref{Cond1}) is fulfilled provided the
inequality

\begin{equation}
\bar{w}<<\frac{v}{2h}\sqrt{\frac{l}{3\xi _{LR}}}\;  \label{wLargeH}
\end{equation}%
is satisfied. Combining Eqs.(\ref{rLargeH}-\ref{wLargeH}), one obtains that
the ratio of the LRTC and singlet component at the SF interface satisfies
the condition

\begin{equation}
r<\sqrt{3\frac{\xi _{LR}}{l}}
\end{equation}

If the spin-dependent scattering can be neglected, this inequality can be
written as $r<(2T\tau )^{-1/4}.$This means that for $T\approx 4K$\ and $\tau
\approx 10^{-14}s$\ the ratio $r$\ should be: $r\lessapprox 6,$\ that is the
amplitude of the LRTC at the SF interface may be comparable with or even
larger than the singlet component.

We see that at a given width of the DW $w$, the amplitude of the LRTC
increases with increasing the exchange field $h$, whereas the amplitude of
the singlet component $s_{3}(0)$ decreases \ and reaches an asymptotic value
$\sim b_{\mu }$ at $h\tau >>1$. The maximum value of the LRTC at $h\tau >>1$
is of the order $\sim b_{\mu }\sqrt{\xi _{LR}/l}.$ The upper limit on $h$ is
imposed by the condition: $w<v/h$, i.e. $\max h\approx v/w.$ In the both
cases of small and large product $h\tau $ the amplitude of the LRTC is
proportional to the width of the DW turning to zero at $w=0$.

\section{Josephson effect}

In this section we consider the dc Josephson effect in an SFS junction with
narrow DWs located at the left and right interfaces. We assume that the
distance between the superconductors is larger than the correlation length $%
\xi _{SN}=\sqrt{D/\Delta }$ in the absence of the exchange field. In this
case, the Josephson coupling is caused only by the LRTC and the overlap of
the LRTC created by each interface is weak. Then, in order to calculate the
Josephson critical current, one can represent the amplitude of the LRTC in
the form

\begin{equation}
\check{s}(x)=\check{s}_{L}(x)+\check{s}_{R}(x)  \label{JEs}
\end{equation}%
where $\check{s}_{L,R}(x)$ are the amplitudes of the LRTC created by the
left (right) interfaces. These matrices are equal to

\begin{eqnarray}
\check{s}_{L}(x) &=&-\hat{\sigma}_{1}\otimes \hat{\tau}_{3}\cdot \hat{f}_{S}%
\frac{\xi _{LR}}{l}B_{L}\exp (-x/\xi _{\omega }),  \label{JEsL} \\
\check{s}_{R}(x) &=&-\hat{\sigma}_{1}\otimes \hat{\tau}_{3}\cdot \hat{S}%
\cdot \hat{f}_{S}\cdot \hat{S}^{\dagger }\frac{\xi _{LR}}{l}B_{R}\exp
(-(L-x)/\xi _{\omega })  \label{JEsR}
\end{eqnarray}
where the coefficients $b_{L,R}$ equal: $B_{L,R}=18\mathit{H}_{\omega
L,R}\langle b_{\mu }\rangle _{L,R}Re(1/K_{h})$ if $h\tau <<1$ and $B_{L,R}=6%
\mathit{H}_{\omega L,R}\langle b_{\mu }/\mu \rangle _{L,R}$ if $h\tau >>1$.

With the help of the matrix $\hat{S}=\cos (\varphi /2)+i\hat{\tau}_{3}\sin
(\varphi /2)$ we take into account the phase difference $\varphi $ between
the superconductors S (the phase of the left S is set to be equal to zero).
In order to compare the magnitude of the Josephson critical current in the
considered case of the SFS junction with the one for an SNS junction, we
write down here also the amplitude of the singlet component for the SNS
junction expressed in terms of the same quantities. We can obtain it from
Eq. (\ref{DifSpm}) simply setting $h=0$ and the corresponding expressions
take the form

\begin{eqnarray}
\check{s}_{L}(x) &=&3\hat{f}_{S}\otimes \hat{\sigma}_{3}\langle b_{\mu
}\rangle \frac{\xi _{LR}}{l}\exp (-x/\xi _{\omega }),  \label{JEsnsL} \\
\check{s}_{R}(x) &=&3\hat{S}\cdot \hat{f}_{S}\otimes \hat{\sigma}_{3}\cdot
\hat{S}^{\dagger }\langle b_{\mu }\rangle \frac{\xi _{LR}}{l}\exp
(-(L-x)/\xi _{\omega }),
\end{eqnarray}%
with $\hat{f}_{S}=f_{S}\hat{\tau}_{2},$ $f_{S}=\Delta /\sqrt{\omega
^{2}+\Delta ^{2}}$.

The current through the SFS (or SNS) junction in the limit $T\tau <<1$ is
given by the expression

\begin{equation}
I=\frac{1}{16}\mathcal{S}\sigma (4\pi Ti)Tr(\hat{\sigma}_{0}\otimes \hat{\tau%
}_{3})\sum_{\omega }\{\check{s}(x)\partial \check{s}(x)/\partial x\}
\end{equation}
$\;$where $\mathcal{S}$ is the cross section area of the junction and the
summation is performed over the fermionic Matsubara frequencies.

Substituting the function $\check{s}$ from Eqs. (\ref{JEs}- \ref{JEsR}) into
this expression we obtain for the case of identical interfaces

\begin{equation}
I_{J}=\frac{9}{8}\mathcal{S}\sigma (4\pi Ti)\langle \gamma \mu \rangle ^{2}Tr%
\hat{\tau}_{3}\sum_{\omega }\{\hat{f}_{S}\cdot \hat{S}\cdot \hat{f}_{S}\cdot
\hat{S}^{\dagger }-\hat{S}\cdot \hat{f}_{S}\cdot \hat{S}^{\dagger }\cdot
\hat{f}_{S}\}\frac{\xi _{LR}}{l}\exp (-L/\xi _{\omega })  \label{JEjosCur1}
\end{equation}%
Calculating the trace in Eq. (\ref{JEjosCur1}), we find

\begin{equation}
I_{J}=I_{c(SNS)}\sin \varphi ,\text{ }I_{c(SNS)}3\sqrt{3/2}\mathcal{S}\sigma
(4\pi T)\langle b_{\mu }\rangle ^{2}\sum_{\omega =0}f_{S}^{2}(\omega )\frac{%
\exp (-L/\xi _{LR})}{l\sqrt{\omega }}  \label{JEcritCur1}
\end{equation}

A similar formula for $I_{J}$ can be obtained for the SFS junction with the
use of the LRTC $\check{s}$ given by Eqs.(\ref{JEsL},\ref{JEsR}). We write
down the expression for the critical currents caused by the LRTC

\begin{equation}
I_{c(SFS)}=-\mathit{H}_{\omega L}\mathit{H}_{\omega R}6\sqrt{6}\mathcal{S}%
\sigma (4\pi T)\langle b_{\mu }\rangle _{L}\langle b_{\mu }\rangle
_{R}\sum_{\omega =0}f_{S}^{2}\frac{\exp (-L/\xi _{\omega })}{l\sqrt{\omega }}%
,\text{ \ }h\tau >>1  \label{e3}
\end{equation}%
where $\xi _{\omega }$\ is given, as before, by Eq. (\ref{e2}).\textbf{\ }

\textbf{\ }The sign opposite to the critical current $I_{c(SNS)}$ in Eq. (%
\ref{e3}) arises because the product $(\hat{f}_{S}\cdot \hat{S}\cdot \hat{f}%
_{S}\cdot \hat{S}^{\dagger })$ in Eq. (\ref{JEjosCur1}) is replaced in the
case of SFS junction by the product $(\hat{\tau}_{3}\cdot \hat{f}_{S}\cdot
\hat{S}\cdot \hat{\tau}_{3}\cdot \hat{f}_{S}\cdot \hat{S}^{\dagger }).$ If
the interfaces and domain walls in SFS are identical ($\mathit{H}_{L}=%
\mathit{H}_{R}\equiv \mathit{H}$), we get
\begin{equation}
I_{c(SFS)}=-4\mathit{H}^{2}I_{c(SNS)}
\end{equation}%
where $\mathit{H}=(h\tau )(w/l)\langle \sin \alpha \rangle _{w}$.

According to the inequality (\ref{Cond1}) the quantity $4\mathit{H}^{2}$
must satisfy the condition $4\mathit{H}^{2}<\sqrt{\omega \tau }<1.$ This
means that the critical current in SFS junctions with a narrow domain wall
is smaller than the critical\textbf{\ }current $I_{c(SNS)}$ of the SNS
junction. However, it can become comparable with the latter provided the
parameter $\mathit{H}$ is of the order of $1$, which is possible for strong
ferromagnets \textbf{(}$h\tau \gg 1)$. In the case of different orientations
of magnetization in the right and left domain walls, i. e., if the product $%
\langle \sin \alpha \rangle _{wL}\langle \sin \alpha \rangle _{wR}$ is
negative, the critical current $I_{c(SFS)}$ has the same sign as $%
I_{c(SNS)}. $ This result is in accordance with the results of Ref. \cite%
{BVE03}, where the sign of the critical current was shown to be sensitive to
a so called chirality depending on whether the magnetization vector $\mathbf{%
M}$ rotated or oscillated when going from one interface to the other. The
negative sign of the critical current in the SFS junction with a half metal
was obtained also in Ref. \cite{Eschrig}.

\section{Spin dependent scattering in a diffusive SF bilayer}

In this section we consider for completeness the diffusive limit assuming
that the mean free path $l$ is shorter than the \textquotedblleft
magnetic\textquotedblright\ length $\xi _{h}$, Eq.(1). However, in contrast
to Ref. \cite{BVE01}\textbf{\ }the width of the DW, $w,$ is supposed to be
shorter than the length $\xi _{h}$. Then, in order to find the LRTC, we can
use the same method as in the preceding sections. We also take into account
the spin dependent scattering that strongly affects the penetration length
of the LRTC. In the case of a diffusive SF bilayer considered here, one can
use the Usadel equation which in the F layer has the form

\begin{equation}
D\partial (\check{g}\partial \check{g}/\partial x^{2})\;-[\omega \hat{\tau}%
_{3}-ih\hat{\tau}_{3}\otimes \hat{\sigma}_{3}\cos \alpha (x),\check{g}]+ih[%
\hat{\tau}_{0}\otimes \hat{\sigma}_{2}\sin \alpha (x),\check{g}]=\check{I}%
_{m}/\tau ,  \label{Usadel}
\end{equation}%
\textbf{\ }where $\check{g}$ is a $4\times 4$ matrix Green's function in the
ferromagnetic region that does not depend in the diffusive limit on the
momentum orientation, $D=vl/3$ is the diffusion coefficient. The matrix in
the R.H.S. is the spin dependent collision term

\begin{equation}
2\check{I}_{m}=\{\check{m}\langle \check{g}\rangle \check{m}\check{g}-\check{%
g}\check{m}\langle \check{g}\rangle \check{m}+\langle \check{g}\rangle _{so}%
\check{g}-\check{g}\langle \check{g}\rangle _{so}\}_{\varphi }.  \label{Im}
\end{equation}%
with $\check{m}=1+\lambda _{z}\check{n}_{z}+\lambda _{\perp }\check{n}%
_{\perp }$, $\check{n}_{z}=\hat{\tau}_{3}\otimes \hat{\sigma}_{3}$, $%
\check{n}_{\perp }=\hat{\tau}_{0}\otimes (\hat{\sigma}_{1}\cos \varphi +\hat{%
\sigma}_{2}\sin \varphi ).$ The subscript $\varphi $ means the averaging
over the azimuthal angle $\varphi $.

The last two terms in Eq. (\ref{Im}) stand for the spin-orbit scattering

\begin{equation}
\langle \check{g}\rangle _{so}=(\lambda _{so}/4\pi )\int d\Omega ^{\prime
}e_{i}^{\prime }e_{k}^{\prime }(\mathbf{S\times e})_{i}\check{g}(\mathbf{%
S\times e})_{k}.
\end{equation}%
with $\mathbf{S=(}\hat{\sigma}_{1},\hat{\sigma}_{2}\mathbf{,\hat{\tau}}%
_{3}\otimes \hat{\sigma}_{3}\mathbf{)}$. The coefficients $\lambda _{z,\perp
}$ and $\lambda _{so}$ are expressed in terms of spatial fluctuations of the
magnetic moments of impurities (see Refs.\cite{BVErmp,BVE07}). For example,
the most important coefficient, $\lambda _{so}$, related to the spin-orbit
interaction is equal to $\lambda _{so}=\tau /\tau _{so}$, $\tau ^{-1}=2\pi
\nu N_{imp}u_{imp}^{2}$, $\tau _{so}^{-1}=2\pi \nu N_{imp}\int d\Omega /4\pi
u_{so}^{2}\sin ^{2}\theta $, where $\nu $ is the density of states at the
Fermi level, which is assumed to be the same for the spin-up and down
orientations in the quasiclassical approximation, $N_{imp}$ is the impurity
concentration, $u_{imp}$ and $u_{so}$ is the potential of impurities and
spin-orbit interaction, respectively. These coefficients are related to the
quantities used in Ref. \cite{BuzdinSO}, $\Gamma _{x,z}$ and $\Gamma _{so},$
in the following way: $2\tau \Gamma _{x,z}=$ $\lambda _{\perp ,z}$ and $%
9\tau \Gamma _{so}=\lambda _{so}$.

We employ the boundary condition at the SF interface in the form presented
in Ref. \cite{KL}

\begin{equation}
\check{g}\partial \check{g}/\partial x|_{x=0}=(2\gamma _{F})^{-1}[\check{g}%
_{S},\check{g}]  \label{DiffBC}
\end{equation}%
where $\gamma _{F}=R_{B}\sigma _{F}=l/b_{av}$, $b_{av}=c_{1}T_{av},$ $T_{av}$
is an effective transmission coefficient averaged over angles, $c_{1}$ is a
numerical factor of the order $1$ \cite{Zaitsev,KL}. Integrating Eq. (\ref%
{Usadel}) over the width of the DW, we obtain an effective boundary
condition for the Usadel equation

\begin{equation}
\check{g}\partial \check{g}/\partial x|_{x=w}=(2\gamma _{F})^{-1}[\check{g}%
_{S},\check{g}]-iK_{D}[\hat{\tau}_{0}\otimes \hat{\sigma}_{2},\check{g}]
\label{DiffEffBC}
\end{equation}
where $K_{D}=(hw/D)\langle \sin \alpha (x)\rangle _{w}\equiv h\bar{w}/D.$

We assume again that the proximity effect is weak so that the matrix $\check{%
g}$ can be represented in the form of Eq. (\ref{gNorm}). Then, we linearize
Eqs.(\ref{Usadel}, \ref{DiffEffBC}) and arrive at the equation for $\check{f}
$ in the region outside the domain wall ($x>w$)

\begin{equation}
\partial ^{2}\check{f}/\partial x^{2}-2\varkappa _{\omega }^{2}\check{f}%
+i\varkappa _{h}^{2}[\hat{\sigma}_{3},\check{f}]_{+}=\varkappa
_{non}^{2}\delta \check{I}_{m}.  \label{UsadelLin}
\end{equation}
where $\varkappa _{non}=1/\sqrt{D\tau }$ is a wave vector related to a
nonmagnetic scattering$,$ $\delta \check{I}_{m}=\delta \check{I}_{sp}+\delta
\check{I}_{so}$ and

\begin{eqnarray}
\delta \check{I}_{sp} &=&\{\lambda _{z}^{2}(\check{f}+\hat{\sigma}%
_{3}\otimes \check{f}\otimes \hat{\sigma}_{3})+\lambda _{\perp }^{2}[\check{f%
}-(\hat{\sigma}_{1}\otimes \check{f}\otimes \hat{\sigma}_{1}+\hat{\sigma}%
_{2}\otimes \check{f}\otimes \hat{\sigma}_{2})/2]\}  \label{Isp} \\
\delta \check{I}_{so} &=&(\lambda _{so}/3)\{\check{f}+(\hat{\sigma}%
_{1}\otimes \check{f}\otimes \hat{\sigma}_{1}+\hat{\sigma}_{2}\otimes \check{%
f}\otimes \hat{\sigma}_{2}-\hat{\sigma}_{3}\otimes \check{f}\otimes \hat{%
\sigma}_{3})/3\}  \label{Iso}
\end{eqnarray}

The effective boundary conditions are obtained as before and have the form

\begin{equation}
\partial \check{f}/\partial x|_{x=0}=(1/\gamma _{F})(|g_{S}|\check{f}-\check{%
f}_{S})-iK_{D}\hat{\tau}_{3}\otimes \lbrack \hat{\sigma}_{2},\check{f}],%
\text{ }\partial \check{f}/\partial x|_{x=d}=0  \label{DiffBClin}
\end{equation}

Here $\varkappa _{\omega }^{2}=|\omega |/D,$ $\varkappa _{h}^{2}=-h_{\omega
}/D,$ $K_{D}=h_{\omega }w\langle \sin \alpha (x)\rangle _{w}/D,$ $%
g_{S}=\omega /\sqrt{\omega ^{2}+\Delta ^{2}}$.

We again seek for a solution in the form

\begin{equation}
\check{f}(x)=\hat{\tau}_{2}\otimes (\hat{\sigma}_{3}f_{3}(x)+\hat{\sigma}%
_{0}f_{0}(x))+\hat{\tau}_{1}\otimes \hat{\sigma}_{1}f_{1}(x)  \label{DiffF}
\end{equation}
where $f_{3}(x)$ is the amplitude of the singlet component and $f_{0,1}(x)$
are the amplitudes of the short-range $S_{z}=0$ and long-range $|S_{z}|=1$
triplet components, respectively.

In this section we consider a SF bilayer of a finite width having in mind to
calculate the DOS variation at the outer surface of the F layer. In order to
satisfy the second boundary condition at $x=d$, Eq. (\ref{DiffBClin}), we
represent the solution in the form

\begin{eqnarray}
f_{0,3}(x) &=&C_{0,3+}\cosh (\varkappa _{+}(x-d))+C_{0,3-}\cosh (\varkappa
_{-}(x-d)),  \label{DiffFx} \\
f_{1}(x) &=&C_{1}\cosh (\varkappa _{1}(x-d))  \label{DiffFx1}
\end{eqnarray}%
with the decay lengths determined by $\varkappa _{\pm }$ and $\varkappa
_{1}. $

Substituting Eqs. (\ref{DiffF}-\ref{DiffFx1}) into Eq. (\ref{UsadelLin}), we
obtain a system of equations for the coefficients $C_{0,3}$ and $C_{1}$

\begin{eqnarray}
C_{0}(\varkappa ^{2}-2\varkappa _{\omega }^{2}-K_{0}^{2})+2i\varkappa
_{h}^{2}C_{3} &=&0,  \label{DiffC0} \\
C_{3}(\varkappa ^{2}-2\varkappa _{\omega }^{2}-K_{3}^{2})+2i\varkappa
_{h}^{2}C_{0} &=&0,  \label{DiffC3} \\
C_{1}(\varkappa ^{2}-2\varkappa _{\omega }^{2}-K_{1}^{2}) &=&0,
\label{DiffC1}
\end{eqnarray}%
where $K_{0}^{2}=2\varkappa _{non}^{2}(\lambda _{z}+(2/9)\lambda _{so}),$ $%
K_{3}^{2}=2\varkappa _{non}^{2}(\lambda _{z}+\lambda _{\perp }),$ and $%
K_{1}^{2}=\varkappa _{non}^{2}(\lambda _{\perp }+(4/9)\lambda _{so}).$

We see that the wave vector characterizing the decay of the singlet
component does not depend on the spin-orbit scattering as it should be. Note
that the influence of the spin-orbit scattering on the SR components has
been considered for the first time in Ref. \cite{Demler}. The Eigenvalue of
Eq.(\ref{DiffC1}) $\varkappa _{1}^{2}$ equals

\begin{equation}
\varkappa _{1}^{2}=2\varkappa _{\omega }^{2}+K_{1}^{2}  \label{DiffK1}
\end{equation}

The Eigenvalues $\varkappa _{\pm }^{2}$ that determine the relation between
the coefficients $C_{0,3}$ are found from Eqs.(\ref{DiffC0}-\ref{DiffC3}).
They are the roots of the equation

\bigskip
\begin{equation}
(\varkappa ^{2}-2\varkappa _{\omega }^{2}-K_{0}^{2})(\varkappa
^{2}-2\varkappa _{\omega }^{2}-K_{3}^{2})+4\varkappa _{h}^{4}=0
\label{DiffRoots}
\end{equation}%
As follows from Eq.(\ref{DiffRoots}), both the Eigenvalues are real provided
the condition

\begin{equation}
4\varkappa _{h}^{2}<|K_{0}^{2}-K_{3}^{2}|,
\end{equation}%
is fulfilled. In this case there are no oscillations in the condensate
functions and, therefore, no oscillations of observable quantities.

In the limit

\begin{equation}
\varkappa _{h}^{2}>>2\varkappa _{\omega }^{2},K_{0,3,1}^{2}  \label{DiffCond}
\end{equation}%
the Eigenvalues equal

\begin{equation}
\varkappa _{\pm }^{2}\approx \pm 2i\varkappa _{h}^{2}+2\varkappa _{\omega
}^{2}+(K_{0}^{2}+K_{3}^{2})/2  \label{DiffKpm}
\end{equation}

The coefficients $C_{0,3}$ and $C_{1}$ are found from Eqs.(\ref{DiffC0}-\ref%
{DiffC3}) and the first boundary condition, Eq. (\ref{DiffBClin}). Under the
condition (\ref{DiffCond}) they are equal to

\begin{equation}
C_{3\pm }\approx \mp C_{0\pm }\approx \frac{d}{\gamma _{F}}\frac{1}{2A_{\pm }%
}f_{s},\text{ }C_{1}\approx \frac{K_{D}d^{2}}{\gamma _{F}}\frac{1}{A_{1}}%
f_{s}(\frac{1}{\tilde{A}_{+}}+\frac{1}{\tilde{A}_{-}})  \label{DiffC031}
\end{equation}%
where $f_{s}=\Delta /\sqrt{\omega ^{2}+\Delta ^{2}}$, $A_{\pm }=\theta _{\pm
}\sinh \theta _{\pm }+(d/\gamma _{F})g_{S}\cosh \theta _{\pm },$ $%
A_{1}=\theta _{1}\sinh \theta _{1}+(d/\gamma _{F})g_{S}\cosh \theta _{1},%
\tilde{A}_{\pm }=\theta _{\pm }\tanh \theta _{\pm }+(d/\gamma
_{F})g_{S},\theta _{\pm }=\kappa _{\pm }d,\theta _{1}=\kappa _{1}d.$ Again,
we neglected the influence of the LRTC on the SR components. This is
justified provided the condition

\begin{equation}
2K_{D}dC_{1}\cosh\theta_{1}<<(d/\gamma _{F})f_{S}
\end{equation}%
is fulfilled.

As follows from Eq. (\ref{DiffK1}), the spin-dependent scattering can
essentially reduce the penetration depth for the LRTC. This holds also for
the cases considered in Secs. II-IV.

Eqs.(\ref{DiffFx}-\ref{DiffC031}) describe the spatial dependence of the SR,
$f_{3,0}(x)$, and LR, $f_{1}(x),$ components. In particular, at the outer
boundary of the ferromagnet we have $f_{3,0}(d)=\pm C_{3+}+C_{3-}$ and $%
f_{1}(d)=C_{1}.$ This means that the short-range components oscillate and
decay over \ a distance of the order of $\xi _{h}:$ $f_{3,0}(d)\sim \exp
(-(1+i)d/\xi _{h})$ at $d/\xi _{h}>>1$, whereas the LRTC, $f_{1}(d)$, decays
in a monotonous way over a longer distance $\sim \varkappa _{1}^{-1}$. The
ratio of the LRTC and singlet component at the interface is equal to

\begin{equation}
r=\frac{|f_{1}(0)|}{|f_{3}(0)|}=\frac{2w(h/D)\langle \sin \alpha \rangle _{w}%
}{\varkappa _{1}\tanh \theta _{1}+\gamma _{F}^{-1}|g_{S}|}
\end{equation}%
This quantity may be both larger or less than unity.

In the next section we calculate the DOS by using the results for $%
f_{0,3,1}(x)$ obtained here.

\section{DOS in a diffusive SF bilayer}

The DOS variation, $\delta \nu =\nu -1,$ in the ferromagnetic film caused by
proximity effect in SF bilayers was measured in a number of works \cite%
{KontosDOS,Cretinon,Beasley}. In particular, the inversion of $\delta \nu $
with increasing the thickness of the F layer $d$ was observed. This effect
has been explained theoretically in terms of the SR component oscillations
in space \cite{NazarovDOS,Cretinon,Valls,TanakaDOS}.

An interesting, although small, effect has been observed in a recent work %
\cite{Beasley}. The authors measured the DOS at the outer surface of the
ferromagnet F in a SF system for various thicknesses of the ferromagnet $d$.
They identified two small peaks in the variation of the DOS. One of these
peaks corresponded to the energy gap $\Delta $ in the superconductor,
whereas the other one corresponded to a smaller energy. The first peak
inverted with increasing $d$ but the sign of the second peak remained
unchanged.

The authors of Ref.\cite{Beasley} suggested an explanation of this effect
assuming that the second peak is due to a contribution of the LRTC. At the
same time, this peak cannot be a result of a long-range penetration of the
LRTC into the ferromagnet but is rather due to a different (monotonous)
dependence on the thickness $d$.

In this section, we represent the contributions of the SR and LR components
to the DOS in the ferromagnetic layer using Eqs.(\ref{DiffFx}-\ref{DiffFx1},%
\ref{DiffC031}). We demonstrate that the contribution due to the SR
components, as it was shown earlier, changes the sign with increasing $d$,
while the contribution due to the LR component does not. We are not going to
make a detailed comparison with the experimental results because not all
necessary data are available. For example, nothing is known about the domain
structure in the F layer.

In calculating the DOS, we use parameters close to estimates presented in
Ref. \cite{Beasley}: $(\gamma _{F}d)^{-1}=d/(\xi _{s}\gamma _{B})\approx 0.3$
for $d=4nm,$ $\gamma _{B}=0.5$ and $\xi _{s}=\sqrt{D/2\Delta }=23nm.$

In the considered case of a weak proximity effect, the correction to the
DOS, $\delta \nu (\epsilon )$, at boundary $x=d$ is equal to

\begin{equation}
\delta \nu (\epsilon
)=-(1/2)Re(f_{3}^{2}(d)+f_{0}^{2}(d)+f_{1}^{2}(d))_{\omega =-i\epsilon }
\end{equation}

\begin{figure}[tbp]
\begin{center}
\includegraphics[width=8cm, height=6cm]{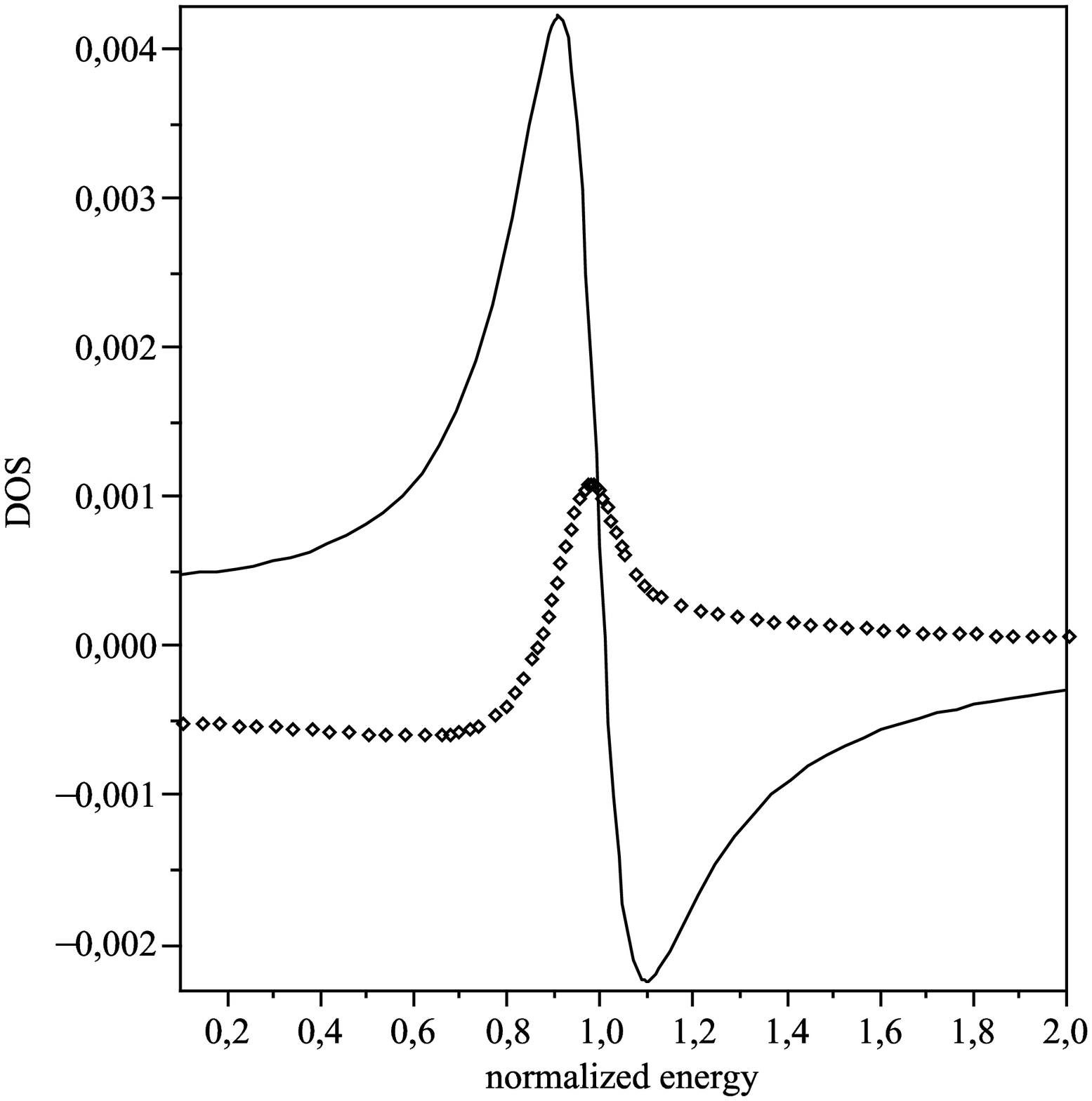} %
\includegraphics[width=8cm, height=6cm]{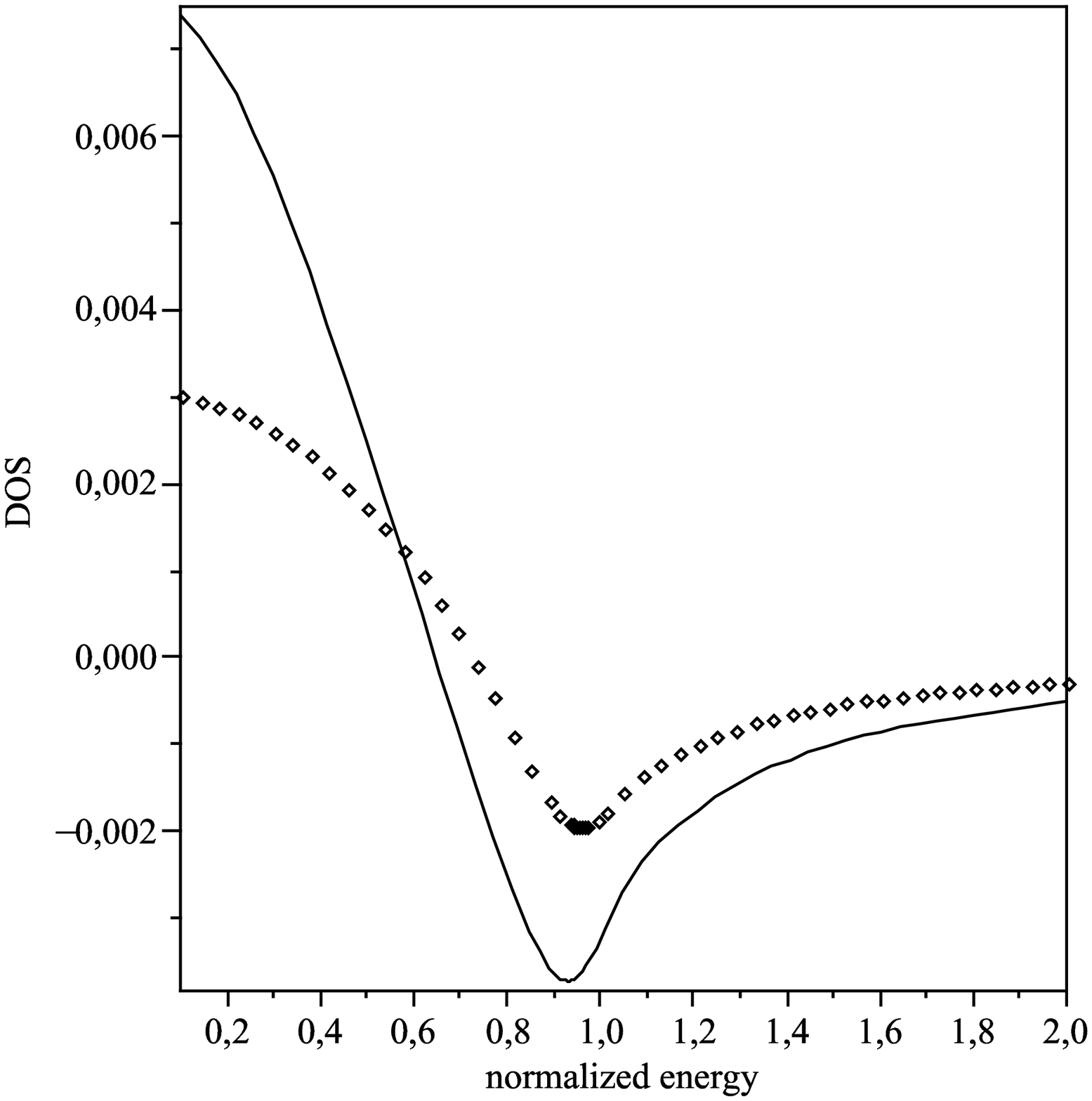}
\end{center}
\caption{DOS variation $\protect\delta \protect\nu (\protect\epsilon )$ at
the outer surface of the F layer vs the normalized energy $\protect\epsilon %
/\Delta $ plotted on the basis of Eqs.(\ref{DiffFx}-\ref{DiffFx1},\ref%
{DiffC031}). The contributions of the SR, $f_{0,3}$, 
 and LRTC, $f_{1}$, components are shown in Fig.2A and Fig.2B, respectively. The
following values of parameters are used: $\protect\theta _{h}\equiv d/%
\protect\xi _{h}=1.5$ (solid lines) and $\protect\theta _{h}=1.8$ (dotted
lines); $\Gamma /\Delta =0.1,$ $d/\protect\gamma _{F}=0.4,$ $%
K_{1}=0.4\varkappa _{h}$. The product $K_{D}\bar{w}d$ is taken to be equal to 0.05.
The parameters $\varkappa _{\protect\epsilon}$ and
$K_{3}$ are assumed to be much smaller than $\varkappa _{h}.$}
\end{figure}

\begin{figure}[tbp]
\begin{center}
\includegraphics[width=8cm, height=6cm]{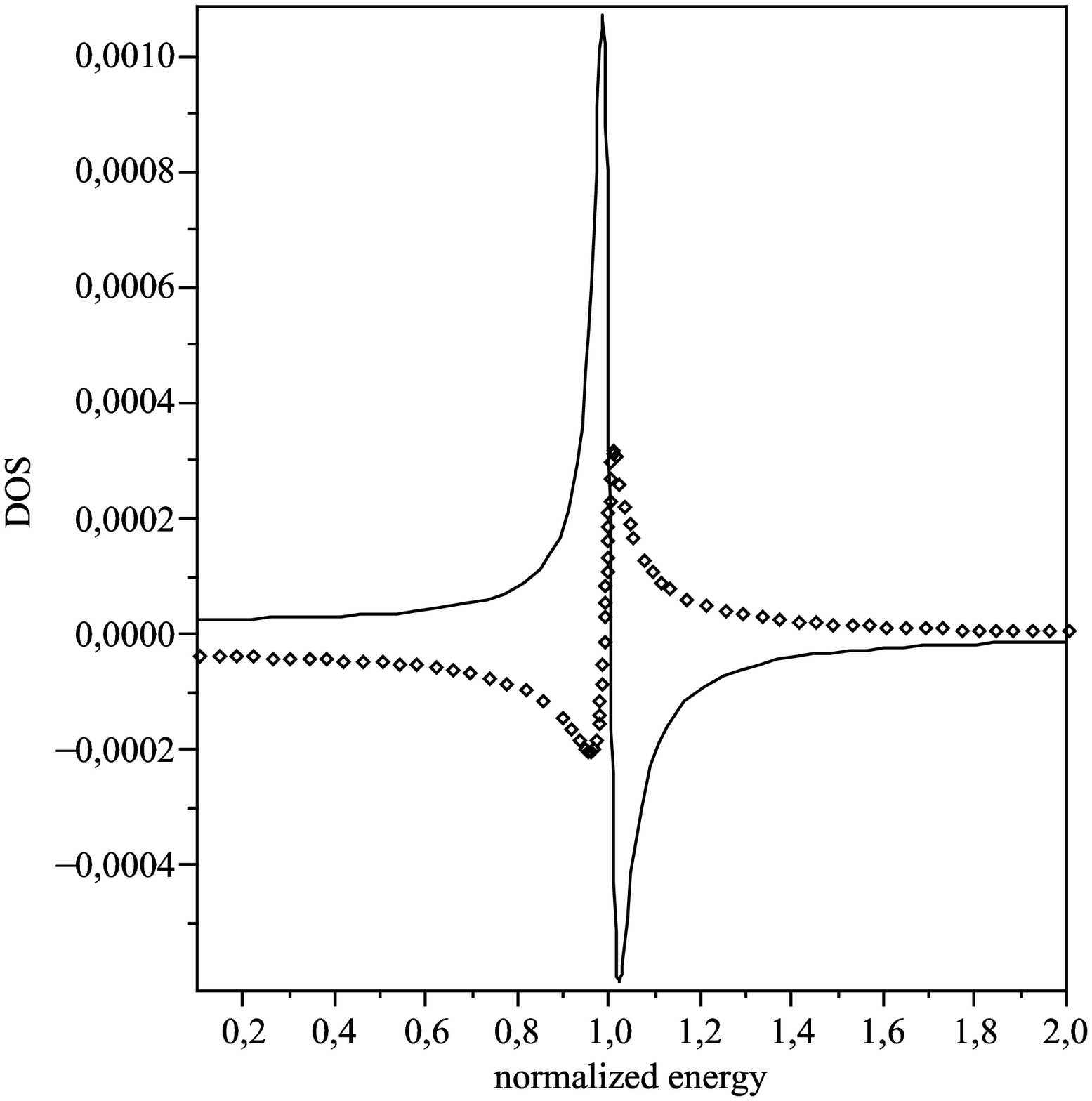} %
\includegraphics[width=8cm, height=6cm]{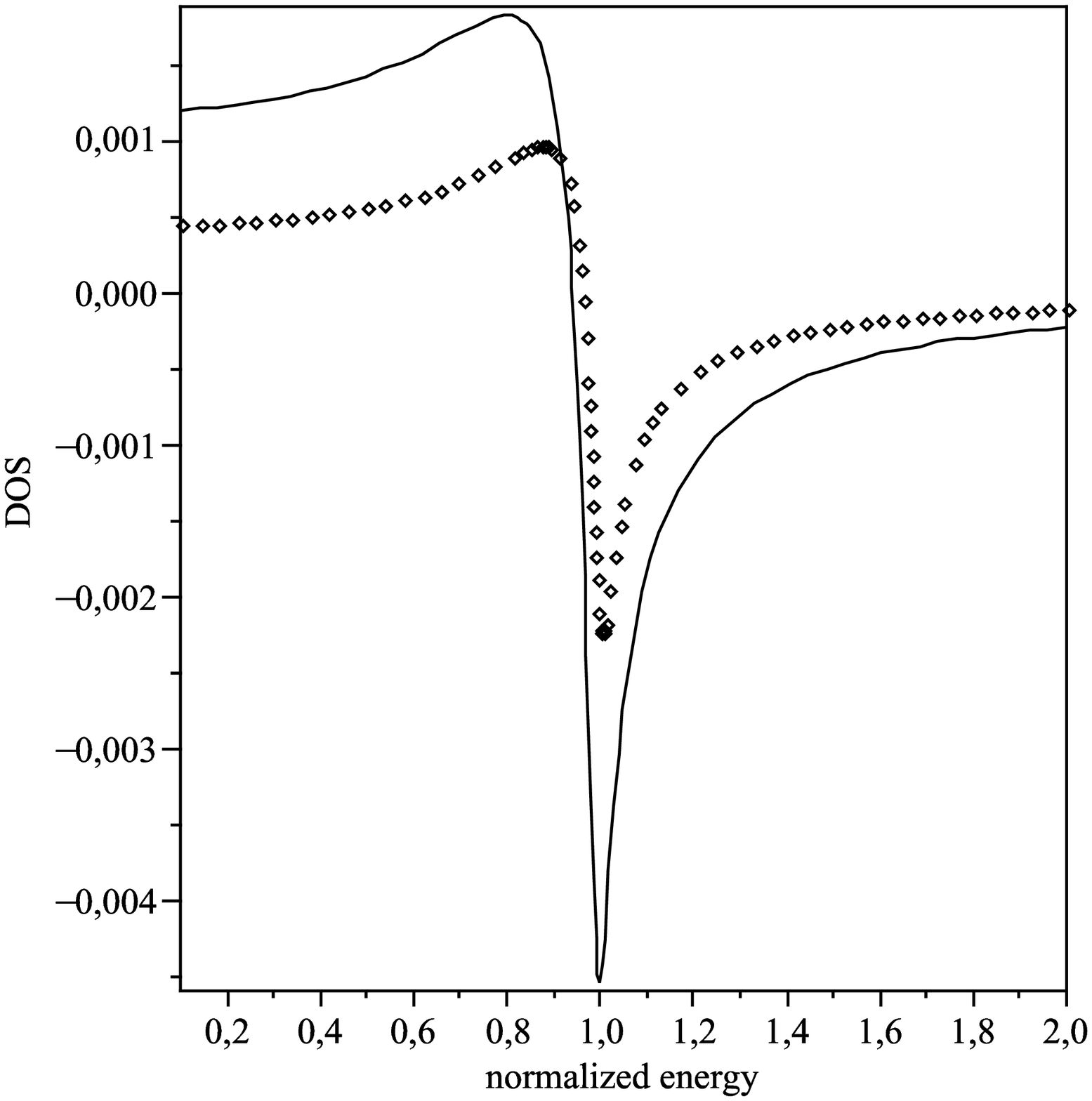}
\end{center}
\caption{The same graphs as in Fig.2 with parameters: $\protect\theta %
_{h}\equiv d/\protect\xi _{h}=1.5$ (solid lines) and $\protect\theta %
_{h}=1.8 $ (dotted lines); $\Gamma /\Delta =0.02,$ $d/\protect\gamma %
_{F}=0.1,$ $K_{1}=0.4\varkappa _{h}$,$d/\protect\gamma _{F}=0.4,$ $%
K_{1}=0.6\varkappa _{h}$. The product $K_{D}\bar{w}d$ is equal to 0.1.
The parameters $\varkappa _{\protect%
\epsilon }$ and $K_{3}$ are assumed to be much smaller than $\varkappa _{h}.$%
}
\end{figure}
where the condensate functions $f_{3,0,1}^{2}(d)$ are determined by Eqs. (%
\ref{DiffFx}-\ref{DiffFx1},\ref{DiffC031}).

In Fig.2 and 3 we plot the contributions to the DOS from the singlet, $f_{3}$%
, SR triplet, $f_{0},$ and LR triplet, $f_{1}$, components as a function of
the energy $\epsilon $ for two different thicknesses of the ferromagnetic
layer. To be more precise, in Figs.2A and 3A the corrections $\delta \nu
_{SR}=-(1/2)Re(f_{3}^{2}(d)+f_{0}^{2}(d))$ due to the SR components are
plotted, whereas in Figs.2B and 3B we show the dependence of $\delta \tilde{%
\nu} _{LR}=a^{-1}\delta \nu _{LR}$ versus energy, where $\delta \nu
_{LR}=-(1/2)Re(f_{1}^{2})$ and $a=(\bar{w}/d)\theta _{h}^{2}$. That is, in
order to get the actual contribution to the DOS due to the LR component the
magnitudes shown in Figs.2B and 3B should be multiplied by $a$.

We see that the corrections to the DOS due to the SR components change sign
with increasing $d$, whereas the sign of the correction due to the LRTC
remains unchanged. It is also worth mentioning that, strictly speaking,
singularities in the SR and LR components correspond to different energies.
If the condition (\ref{DiffCond}) is fulfilled, only the function $%
f_{S}(\epsilon )=i\Delta /\sqrt{(\epsilon +i\Gamma )^{2}-\Delta ^{2}}$
depends on the energy $\epsilon ,$ and therefore the position of
singularities is determined only by the energy gap $\Delta $ and damping
constant $\Gamma .$

On the other hand, the first term in the expression for $A_{1}=\theta
_{1}\sinh \theta _{1}+(d/\gamma _{F})g_{S}(\epsilon )\cosh \theta _{1}$ may
be comparable with the second one that also depends on the energy $\epsilon
. $ The account for the second term leads to a decrease of a characteristic
energy that determines the position of the singularity. One can see that the
contribution of the LRTC is comparable with that of the SR components if the
width of the DW, $w$, is comparable with $\xi _{h}$.

\section{Conclusions}

We have considered the long-range triplet component in an SF bilayer arising
due to an nonhomogeneous magnetization in the F layer (for example, due to a
domain wall) located in the vicinity of the SF interface. Unlike Refs.\cite%
{BVE01,BVErmp} where the width of the DW, $w,$ was assumed to be larger than
the mean free path, we have calculated in the present paper the amplitudes
of the LR as well as of the SR components for the case of a narrow DW. In
fact, our model may be considered as a microscopic model of a spin-active SF
interface usually described by introducing phenomenological parameters.

Assuming that the proximity effect is weak (this corresponds to experimental
data), we have obtained analytical formulas for the amplitudes of the LR and
SR components in a wide range of parameters. The amplitudes of the SR
components decrease with increasing the exchange energy $h$ and become
constant at $h\tau >>1$. The amplitude of the LRTC essentially depends on
the parameter $h\tau $ and increases with increasing $h$. The maximum value
of the amplitude of the LRTC in our approach is determined by the condition $%
h<(v/w)$.

We have calculated the critical Josephson current $I_{c}$ in a SFS junction
where the Josephson coupling is due to the LRTC. The current $I_{c}$ is
negative if the rotation of the magnetization vector $\mathbf{M}$ in DWs at
each SF interface occurs in one direction (positive chirality) and is
positive if the rotation of $\mathbf{M}$ occurs in different direction
(negative chirality).

We have also found the DOS at the outer surface of the F layer in an SF
structure in the presence of a DW at the SF interface. It has been shown
that contributions to the DOS from the SR and LR components have
singularities at an energy $\sim \Delta .$ Whereas the singularity due to
the SR components changes sign with increasing the thickness of the F layer,
$d$, the singularity due to the LR component does not. The change of sign
occurs at $d\approx (\pi /2)\xi _{h}$. Note also that the contribution of
the LRTC to the DOS is of the same order as the one of the SR components
provided the width of the DW is comparable with the length $\xi _{h}$.

We considered the case of the DW parallel to the SF interface. However, this
fact is not crucial: the LRTC created by DWs perpendicular to the SF
interface may be of the same order as the LRTC induced by the DW parallel to
the SF interface. The amplitude of the LRTC for the case of the Neel DWs
perpendicular to the SF interface has been calculated in Ref.\cite{Fominov}.
One can show that similar results can be obtained for the case of the Bloch
DWs perpendicular to the SF interface \cite{Unpubl}. In order to carry out a
more detailed comparison with experiments, more data are required. In
particular, one has to know the parameters of the magnetic structure of the
F film.

\section{Acknowledgements}

We thank SFB 491 for financial support.

\bigskip


\bigskip

\begin{thebibliography}{99}
\bibitem{BCS} J. Bardeen, L.N. Cooper, and J.R. Schrieffer, Phys. Rev. 108,
1175 (1957).

\bibitem{Schrieffer} J.R. Schrieffer, \textit{Superconductivity} (Benjamin,
New York), (1964).

\bibitem{Kirtley} C.C. Tsuei, and J.R. Kirtley, Rev. Mod. Phys. \textbf{72},
969 (2000).

\bibitem{Mineev} V.P.Mineev and K.V.Samokhin, \textit{Introduction to
Unconventional Superconductivity} (Gordon and Breach, Amsterdam; 1999).

\bibitem{Maeno} A.P. Mackenzie and Y. Maeno, Rev. Mod. Phys. \textbf{75},
657 (2003).

\bibitem{Eremin} I. Eremin, D. Manske, S.G. Ovchinnikov, and J. F. Annett,
Ann. Phys. (Berlin) \textbf{13}, 149 (2004).

\bibitem{Berez} V.L. Berezinskii, JETP\ Lett. \textbf{20}, 287 (1975).

\bibitem{Leggett} A.J. Legget, Rev. Mod. Phys. \textbf{77}, 935 (2005).

\bibitem{Wolfle} D.Vollhardt and P.W\"{o}lfle, \textit{The superfluid phases
of He}$^{3}$ (Taylor and Francis. London, New York, Philadelphia; 1990).

\bibitem{BVE01} F.S. Bergeret, A.F. Volkov, K.B. Efetov, Phys. Rev.Lett.
\textbf{78}, 4096 (2001).

\bibitem{BVErmp} F.S. Bergeret, A.F. Volkov, K.B. Efetov, Rev. Mod. Phys.
\textbf{77, }1321 (2005).

\bibitem{BuzdinRMP} A. Buzdin, Rev. Mod. Phys. \textbf{77}, 935 (2005).

\bibitem{AVZ} S.N. Artemenko, A.F. Volkov, and A.V. Zaitsev, Sol. St. Comm.
\textbf{30}, 771 (1979).

\bibitem{GubMar} V.N. Gubankov and N.M. Margolin, JETP\ Lett. \textbf{29},
673 (1979).

\bibitem{Naz} Y.V. Nazarov and T.H. Stoof, Phys. Rev. Lett. \textbf{76}, 823
(1996).

\bibitem{LamVol} A.F. Volkov, N. Allsopp, and C.J. Lambert, J. Phys. Condens
Matter \textbf{8}, 45 (1996).

\bibitem{Pann} P. Charlat, H. Courtois, P. Gandit, D. Mailly, A.F. Volkov,
and B. Pannetier, Phys. Rev. Lett. \textbf{77}, 4950 (1996).

\bibitem{Kadig} A. Kadigrobov, R.I. Shekhter and M. Jonson, Europhys. Lett.
\textbf{54}, 394 (2005).

\bibitem{Eschrig} M. Eschrig, J. Kopu, J.C. Cuevas, and G. Sch\"{o}n, Phys.
Rev.Lett. \textbf{90}, 137003 (2003).

\bibitem{Eschrig08} M. Eschrig, T. Lofwander, Nature Physics \textbf{4},
138-143 (2008).

\bibitem{Tanaka} Y. Asano, Y. Sawa, Y. Tanaka, and Alexander A. Golubov,
Phys. Rev. B \textbf{76}, 224525 (2007).

\bibitem{Zaikin} A.V. Galaktionov, M.S. Kalenkov, and A.D. Zaikin, Phys.
Rev. B \textbf{77}, 094520 (2008)

\bibitem{Eilenberger} G. Eilenberger, Z. Phys. \textbf{214}, 195 (1968).

\bibitem{LO} A.I. Larkin and Yu.N. Ovchinnikov, in \textit{Nonequilibrium
Superconductivity}, edited by D.N. Langenberg and A.I. Larkin (Elsevier,
Amsterdam, 1984).

\bibitem{Keiser} R.S. Keizer, S.T.B. Goennenwein, T.M. Klapwijk, G. Miao, G.
Xiao, A. Gupta, Nature \textbf{439}, 825 (2006).

\bibitem{Sosnin} I. Sosnin, H. Cho, V.T. Petrashov, and A.F. Volkov, Phys.
Rev. Lett. \textbf{96,} 157002 (2006).

\bibitem{Lawrence} M.D. Lawrence and N. Giordano, J. Phys. Cond. Matter, 8,
563 (1996).

\bibitem{Giraud} M. Giraud, H. Courtois, K. Hasselbach, D. Mailly, and B.
Pannetier, Phys. Rev. B \textbf{58}, 11872 (1998).

\bibitem{Petrashov99} V.T. Petrashov, I.A. Sosnin, I. Cox, A. Parsons, and
C. Troadec, Phys. Rev. Lett. \textbf{83}, 3281 (1999).

\bibitem{Chandrasekhar01} J. Aumentado and V. Chandrasekhar, Phys. Rev. B
\textbf{64}, 054505 (2001).

\bibitem{Ryaz} V.V.~Ryazanov, V.A. Oboznov, A.Yu. Rusanov, A.V.
Veretennikov, A.A. Golubov, and J. Aarts, Phys. Rev. Lett. \textbf{86}, 2427
(2001).

\bibitem{Kontos} T.~Kontos, M. Aprili, J. Lesueur, and X. Grison, Phys. Rev.
Lett. \textbf{89}, 137007 (2002).

\bibitem{Blum} Y.~Blum, A. Tsukernik, M. Karpovski, and A. Palevski, Phys.
Rev. Lett. \textbf{89}, 187004 (2002).

\bibitem{Bauer} A. Bauer, J. Bentner, M. Aprili, M.L. Della Rocca, M.
Reinwald, W. Wegscheider, and C. Strunk, Phys. Rev. Lett. \textbf{92},
217001 (2004).

\bibitem{BVE03} A.F. Volkov, F.S. Bergeret, K. B. Efetov, Phys. Rev.Lett.
\textbf{90}, 117006 (2003).

\bibitem{Buzdin07} M. Houzet and A.I. Buzdin, Phys. Rev. B \textbf{76},
060504 (2007).

\bibitem{NazBr} V. Braude and Yu.V. Nazarov, Phys. Rev. Lett. \textbf{98},
077003 (2007).

\bibitem{Sudbo} J. Linder and A. Sudb\o , Phys. Rev. B 76, 064524 (2007).

\bibitem{Fominov} A.F. Volkov, Ya.V. Fominov, and K.B. Efetov, Phys. Rev. B
\textbf{72}, 184504 (2005); Ya.V. Fominov, A.F. Volkov, and K.B. Efetov,
Phys. Rev. B \textbf{75}, 104509 (2007).

\bibitem{Anishchanka06} A.F. Volkov, A. Anishchanka, and K.B. Efetov, Phys.
Rev. B \textbf{73}, 104412 (2006).

\bibitem{Champel08} T. Champel, T. L\"{o}fwander, M. Eschrig, Phys. Rev.
Lett. \textbf{100}, 077003 (2008).

\bibitem{Beasley} P. SanGiorgio, S. Reymond, M. R. Beasley, J.H. Kwon, K.
Char, cond-mat/0712.1322.

\bibitem{Kopnin} N.B. Kopnin, Theory of Nonequilibrium Superconductivity
(Clarendon Press, Oxford, UK, 2001).

\bibitem{Zaitsev} A.V. Zaitsev, JETP 59, 1015 (1984).

\bibitem{GolubovRMP} A.A. Golubov, M.Y. Kupriyanov, and E.Il'ichev, Rev.
Mod. Phys. \textbf{76, }411 (2004)

\bibitem{BVECrCur} F.S. Bergeret, A.F. Volkov, K.B. Efetov, Phys. Rev. B
\textbf{64}, 134506 (2001).

\bibitem{BB} I. Baladie and A. Buzdin, Phys. Rev. B \textbf{64}, 224514
(2001).

\bibitem{FVE} Ya.V. Fominov, A.F. Volkov and K.B. Efetov, Phys. Rev. B
\textbf{72,} 184504 (2005); ibid \textbf{75,} 104509 (2007).

\bibitem{KL} M. Yu. Kupriyanov and V. F. Lukichev, JETP\textbf{\ 67}, 1163
(1988).

\bibitem{BVE02} F.S. Bergeret, A.F. Volkov, K.B. Efetov, Phys. Rev. B
\textbf{65}, 134505 (2002).

\bibitem{BVE07} F.S. Bergeret, A.F. Volkov, K.B. Efetov, Phys. Rev. \textbf{%
75}, 184510 (2007).

\bibitem{BuzdinSO} D. Y. Gusakova, A. A. Golubov, M. Y. Kupriyanov, and A.
Buzdin, JETP Lett. \textbf{83}, 327 (2006).

\bibitem{Demler} E. A. Demler, G. B. Arnold, and M. R. Beasley, Phys. Rev. B
\textbf{55,} 15174 (1997).

\bibitem{KontosDOS} T.~Kontos, M. Aprili, J. Lesueur, and X. Grison, Phys.
Rev. Lett. \textbf{86}, 304 (2001).

\bibitem{Cretinon} L. Cretinon, A. K. Gupta, H. Sellier, F. Lefloch, M.
Faure, A. Buzdin, and H. Courtois, Phys. Rev. B \textbf{72,} 024511 (2005).

\bibitem{NazarovDOS} M. Zareyan, W. Belzig, and Yu. V. Nazarov, Phys. Rev.
Lett. \textbf{86}, 308 (2001).

\bibitem{Valls} K. Halterman, O. T. Valls, Physica C \textbf{420} 111 (2005).

\bibitem{TanakaDOS} T. Yokoyama, Y. Tanaka, Phys. Rev. B \textbf{75}, 132503
(2007).

\bibitem{Unpubl} A. F. Volkov and K. B. Efetov, unpublished.
\end{thebibliography}
\end{document}